\documentstyle[epsf]{l-aa}
\begin{document}
\thesaurus{08(08.16.7; 13.07.2)}
\title{EGRET Observations of Pulsars}
\author{P.~L.~Nolan \inst{1,9}
\and J.~M.~Fierro \inst{1}
\and Y.~C.~Lin \inst{1}
\and P.~F.~Michelson \inst{1}
\and D.~L.~Bertsch \inst{2}
\and B.~L.~Dingus \inst{2,6}
\and J.~A.~Esposito \inst{2,6}
\and C.~E.~Fichtel \inst{2}
\and R.~C.~Hartman \inst{2}
\and S.~D.~Hunter \inst{2}
\and C.~von~Montigny \inst{2,7}
\and R.~Mukherjee \inst{2,6}
\and P.~V.~Ramanamurthy \inst{2,8}
\and D.~J.~Thompson \inst{2}
\and D.~A.~Kniffen \inst{3}
\and E.~Schneid \inst{4}
\and G.~Kanbach \inst{5}
\and H.~A.~Mayer-Hasselwander \inst{5}
\and M.~Merck \inst{5}
}
\offprints{P. L. Nolan}
\institute{
Hansen Experimental Physics Laboratory,
Stanford University,
Stanford CA 94305, USA
\and
Code 662,
NASA/Goddard Space Flight Center,
Greenbelt MD 20771, USA
\and 
Department of Physics,
Hampden-Sydney College,
Hampden-Sydney VA 23943, USA
\and 
Northrop-Grumman Corporation,
Bethpage NY 11714, USA
\and
Max-Planck Institut f\"ur Extraterrestrische Physik,
Giessenbachstrasse,
D-85748 Garching, Germany
\and
USRA Research Associate
\and
NAS-NRC Research Associate
\and
NAS-NRC Senior Research Associate
\and
e-mail: pln@egret1.Stanford.EDU
} 
\date{Received xxxxxxxxxxx; accepted xxxxxxxxx}
\maketitle

\begin{abstract}
EGRET can now study six pulsars since the recent detection of
PSR B1951+32.
Careful analysis of the three brightest pulsars shows that gamma
rays are emitted through essentially the entire rotation.
This calls into question the previous interpretation of the Crab's
low-level emission as due entirely to its nebula.
The pulse light curves have either well-separated double peaks
or very broad peaks which might be resolved into multiple
components.
The spectra are consistent with power laws over a large portion of the
EGRET energy range, and many show a dropoff at a few GeV.
\keywords{Stars: pulsars: individual -- Gamma rays: observations}
\end{abstract}

\section{Introduction}
EGRET continues to observe rotation-powered pulsars in the 30 MeV
-- 20 GeV gamma-ray band.
So far six have been positively detected.
The most recent addition to the list is PSR B1951+32, which is discussed
by Ramanamurthy et al. \cite{murthy:1951}
(see also Brazier et al. \cite{braz:limit}, Ramanamurthy et al. 
\cite{murthy:disc}).
It joins the Crab, Vela, Geminga, PSR B1706$-$44 and PSR B1055$-$52.
Upper limits for emission from  many other pulsars have been
published (Thompson et al. \cite{thom:survey}; Fierro et al.
\cite{fierro:survey}).

\begin{table}[t]
\centering
\caption{Parameters of the High-Energy Gamma-Ray Pulsars}
\bigskip
\begin{tabular}{lcccc}\hline\hline
 & & & & \\[-0.10in]
       & $P$  & $D$  & $B_s$         & $\dot E$ \\
Pulsar & (ms) & (km) & ($10^{12}$~G) & (ergs~s$^{-1}$) \\[0.04in]
\hline
 & & & & \\[-0.10in]
Crab         &  33.4 & 2.0 & 3.7 & $4.5 \times 10^{38}$ \\
Geminga      & 237.1 & 0.25 & 1.6 & $3.3 \times 10^{34}$ \\
Vela         &  89.3 & 0.50 & 3.3 & $6.9 \times 10^{36}$ \\
PSR B1055$-$52 & 197.1 & 1.5 & 1.1 & $3.0 \times 10^{34}$ \\
PSR B1706$-$44 & 102.4 & 1.8 & 3.1 & $3.4 \times 10^{36}$ \\
PSR B1951+32 &  39.5 & 2.5 & 0.48 & $3.7 \times 10^{36}$ \\
\hline
\hline
\end{tabular}

\label{parameters}
\end{table}

Table~\ref{parameters} lists the parameters of the detected high-energy
gamma-ray pulsars, including the spin period $P$, dispersion measure distance
$D$, inferred surface magnetic field $B_s$, and spin-down energy loss rate
$\dot E$.  In general, the pulsars that are detected are the ones that
should be expected.  One can construct a list of pulsars ordered by $\dot
E/D^2$ (Fierro \cite{fierro:diss}).  Five of the six detected pulsars are
found at the top of this list, along with PSR B1509$-$58, which is detected
at $\sim$1 MeV.  Only PSR B1055$-$52 appears much farther down the list.

There are three sources in the second EGRET catalog (Thompson et al. 
\cite{thom:cat2})
which are coincident with the positions of pulsars high on the
$\dot{E}/d^2$ list, PSR B1046$-$58, PSR B1823$-$13 and PSR B1853+01.
No pulsation at the proper period has yet been detected from any of these.
All three are rather noisy pulsars, so their ephemerides are not
as accurate as might be desired (Kaspi et al. \cite{kaspi}; 
Lyne et al. \cite{lyne}).
Also they are in regions of low EGRET exposure and fairly high 
gamma-ray background.

\section{Continuous emission}

Figure 1 shows the light curves for $>$100 MeV gamma rays from the
three brightest pulsars, Crab, Vela, and Geminga.
A logarithmic scale is used to allow the low flux values to be
visible.
To produce these curves, spatial photon maps were made for each of 
twenty equal-width bins in pulsar phase.
For each phase interval, the source flux was found using a
maximum-likelihood technique which searches for a point-source excess
above a structured background (Mattox et al. \cite{mattox:like}).
This method removes any contribution from the diffuse emission near
the pulsar's position.
This should be compared with the older method (Nolan et al. 
\cite{nolan:crab}; Kanbach et al. \cite{kanb:vela}; Bertsch et al.
\cite{bertsch:gem}; Mayer-Hasselwander et al. \cite{hrm:gem}), 
in which the total number of photons is counted in each phase
bin; the lowest value is assumed to be the background contribution,
which can then be subtracted from the others.
The older method implicitly assumes that the pulsar shuts off its 
gamma-ray emission for a substantial portion of each rotation.

\begin{figure}
\epsfxsize=8.8cm \epsfbox{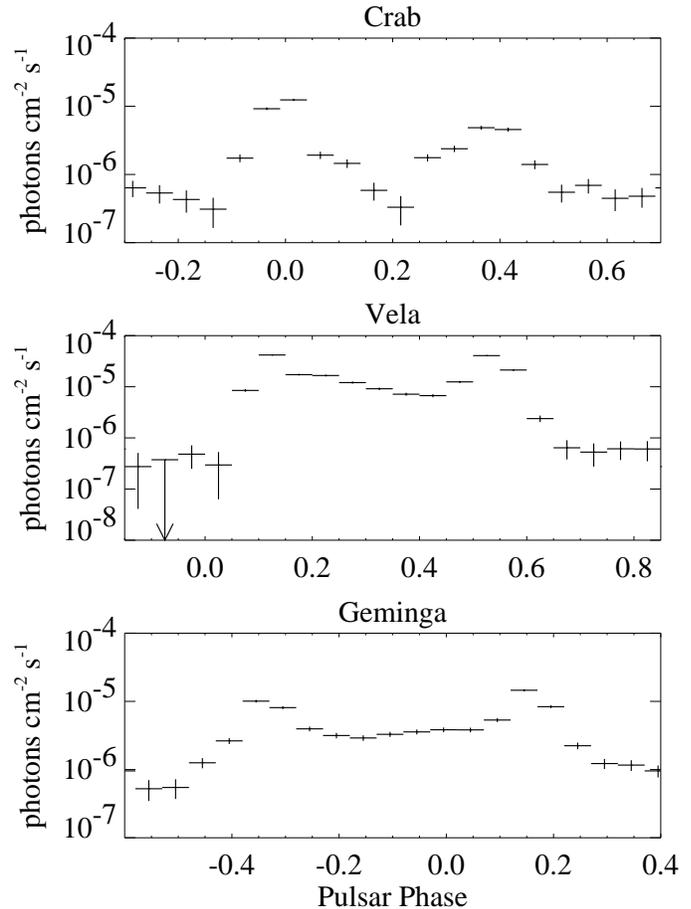}
\caption[]{Light curves of Vela, Geminga and Crab pulsars in $> 100$ MeV
gamma rays.  The background contribution is subtracted from each point, so
this is the true instantaneous source flux.  The Crab and Vela main radio
pulses occur at phase 0.  The Geminga phase definition is arbitrary.}
\end{figure}

The Crab light curve shows the well-known two peaks and a lower level
of emission throughout the rotation.
Note that the flux never drops to zero.
It has been traditional to assume that the weak off-pulse emission between
phase 0.5 and 0.9 is due solely to the Crab Nebula, which EGRET cannot
spatially resolve from the pulsar (Nolan et al. \cite{nolan:crab}; Clear et
al. \cite{clear}).
This tradition can be attributed to an analogy with the behavior
in the X-ray band, where the pulsar and nebula can be spatially
resolved (Harnden \& Seward \cite{harnden}).  
The X-ray pulsar emission drops to nearly zero for much of each rotation
while the nebula continues to shine brightly.

Vela, the next youngest gamma-ray pulsar, shows a similar pattern.
Unlike the Crab pulsar, however, the bridge emission between the two peaks
remains at a very high level.
Still, there is detectable emission through at least 95\% of the rotation.
In previous analysis (Kanbach et al. \cite{kanb:vela}) it has been assumed that
the pulsar shuts off for about 40\% of each rotation.
The detected emission in the off-pulse phase interval is sufficiently weak that
the conclusions of the analysis are still valid.

Geminga (Mayer-Hasselwander et al. \cite{hrm:gem}) shows the same pattern.
It emits through the entire rotation, with strong bridge emission, and
an off-pulse emission level above that from the Crab.
This is troubling because Geminga is over $10^5$ years
old, much too old to have a dense synchrotron/Compton nebula like
the Crab's.

If that is the case, could not the same be true for the Crab?
The belief in the Crab's nebular emission must be called into doubt.
Since the Crab Nebula is certainly there, it must emit gamma rays
at some level.
The best we can say is that the detected off-pulse gamma flux is
an upper limit on the nebular emission.

\section{Average spectra}

The average spectra of the six EGRET-detected pulsars are shown
in Figure 2.
In this analysis, the spectra are derived by the maximum-likelihood
spatial analysis.
In detail these spectra are slightly different from those derived
from the pulse/off-pulse subtraction method.
The gross features, though, are unchanged.

\begin{figure}
\epsfxsize=8.8cm \epsfbox{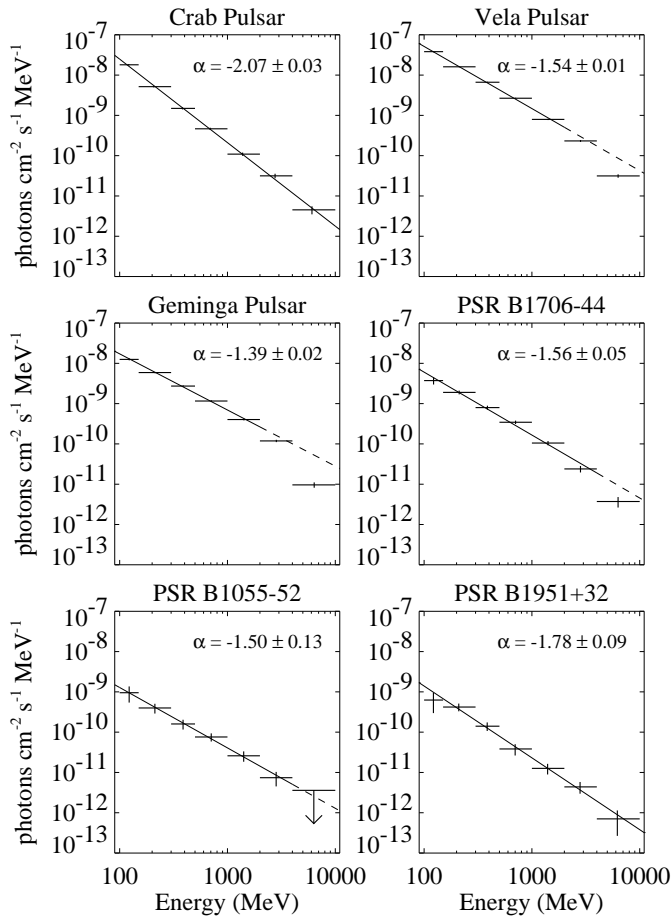}
\caption{Average photon number spectra of the six pulsars detected
by EGRET.  Each can be fitted by a power law over part of the energy
range.}
\end{figure}

Over at least part of the EGRET energy range, each spectrum can
be described quite well by a simple power law.  
The fitted (number) spectral indices range from $-2.07$ for the Crab
to $-1.39$ for Geminga.
With additional data available, the index for PSR B1055-52 is found
to be $-1.50 \pm 0.13$, softer than originally reported, but still
quite hard.
Indeed, four of the six spectra have indices harder than $-1.6$, which may be
significant for emission models.

Three of the spectra show clear evidence for a drop in flux at high
energy, above a few GeV.
Two others have large uncertainties in the 4--10 GeV band, and cannot exclude
the existence of a similar dropoff.
Only the Crab is inconsistent with a significant dropoff above 4 GeV.
Nolan et al. (\cite{nolan:crab}) attributed much of the strong
emission at high energy to the nebula.
As mentioned above, this interpretation is perhaps called into doubt.

A spectral turnover could be due the attenuation of high-energy gamma rays by
the strong magnetic field of the pulsar, or it could be the result of a
dropoff in the charged particle distribution at high energies.  It is
interesting that the gamma ray pulsar with the strongest inferred magnetic
field, Crab, shows no evidence of a spectral turnover, while Geminga, which
has a much weaker inferred magnetic field, has the sharpest spectral
turnover.  The maximum potential drop of a pulsar is estimated to go as
$\Delta\Phi \propto \dot E^{1/2}$ (Sturrock \cite{sturrock}).  Assuming the
maximum energy attainable by a charged particle is directly related to
$\Delta\Phi$, the charged particle distribution of Geminga should turn over
at much lower energies than that of the Crab pulsar, and might explain why
the spectral turnover of Geminga in Figure 2 is more severe than that of the
Crab pulsar.

\section{Light curve shapes}

The gamma-ray light curves of the three weaker pulsars are shown
in Figure 3.
It is possible to draw some general conclusions from these curves
and from Figure 1.

\begin{figure}
\epsfxsize=8.8cm \epsfbox{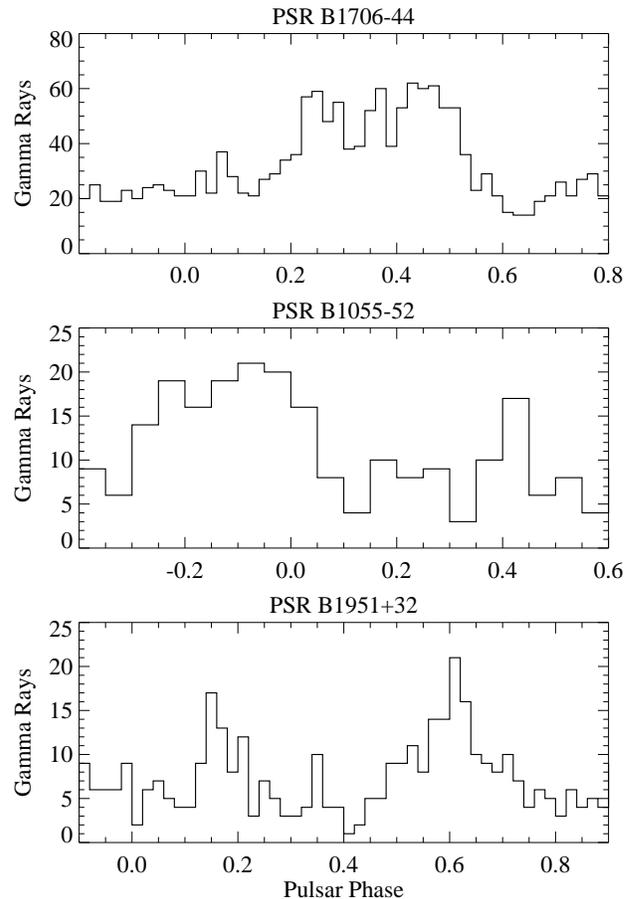}
\caption{Light curves of the three weaker pulsars.  The main radio
pulse occurs at phase 0.}
\end{figure}

The most obvious feature of the light curves is that four of them
have two roughly equal peaks separated by 0.4--0.5 in phase.
The other two, PSR B1055$-$52 (Fierro et al. \cite{fierro:1055}) and 
PSR B1706$-$44 (Thompson et al. \cite{thom:1706}), have broad pulses that might
really be double or even triple.
The suggested narrow peak at phase 0.4 in PSR B1055$-$52 is not
statistically significant.
The apparent triple feature of PSR B1706$-$44 is persistent if the data set
is divided in half according to time or photon energy, but it is still
too early to determine whether this feature is real.
Only improved counting statistics will improve our understanding of
these two pulsars.

Double peaks and broad pulses are fairly rare in radio light curves.  Of
these six pulsars, only the Crab has a gamma-ray light curve which matches
fairly well with the radio pulsation, although the radio emission has a third
peak apparent below 600 MHz that doesn't match any gamma-ray feature.
PSR B1055$-$52 has a similar triple-peaked radio pulse, but there does not
appear to be a correlation between any of the radio and gamma-ray peaks.
The three other radio pulsars have single radio peaks which don't line up
with the gamma-ray peaks.

With the limited statistics available, it is impossible to detect
any emission between the two peaks of PSR B1951+32.
The two peaks are separated by very nearly 0.5 in phase, so
it remains unknown which peak should be called the leading and
which is the trailing.

\section{Trends}

In an attempt to better understand the mechanism responsible for gamma-ray
pulsation, it is natural to examine possible relationships between the
measured gamma-ray characteristics and the inferred pulsar parameters.
With the detection of a sixth high-energy gamma-ray pulsar, as well as
additional exposure to the other pulsars, some of the trends suggested by
early EGRET results stand to be revisited.

One of the more promising trends established by the first five EGRET pulsar
detections was an apparent increase in gamma-ray efficiency with pulsar
characteristic age (Thompson et al. \cite{thom:survey}).  The gamma-ray
efficiency of a
pulsar is defined as the fraction of its available spin-down power which is
emitted in the form of high-energy gamma rays.  Figure 4 shows an updated
plot of the gamma-ray efficiency versus characteristic age, where it has been
arbitrarily assumed that the gamma rays are beamed into a solid angle of 1.0
steradian.  Based on the original five EGRET pulsars, it does indeed seem that
older pulsars convert more of their luminosity into gamma rays.  However, the
gamma-ray efficiency of PSR B1951+32 deviates by an order of magnitude from
the power law trend established by the other pulsars, far more than can be
reasonably attributed to differences in pulsar beaming angles.  Only the Crab
has a measured efficiency which is below that of PSR B1951+32.  Thus, the
presumption that gamma-ray efficiency increases with characteristic age would
appear not to hold true in all cases.

\begin{figure}
\epsfxsize=8.8cm \epsfbox{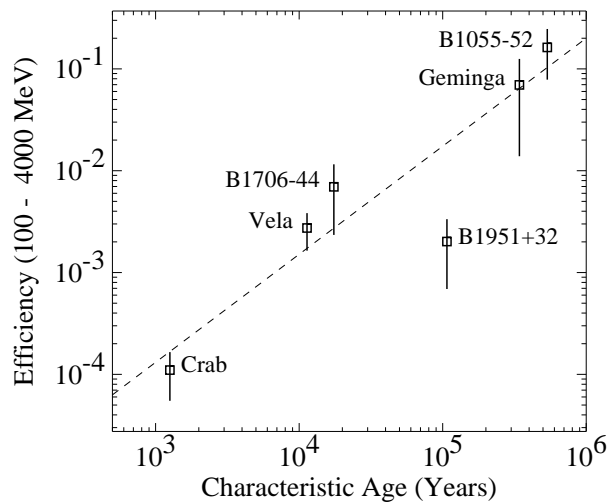}
\caption{Gamma ray production efficiency vs. age.  A beaming angle
of 1.0 steradian has been assumed.}
\end{figure}

Thompson et al. (\cite{thom:survey}) also noted that among the original five
EGRET pulsars, the derived photon spectral indices were harder for the older
pulsars.  However, additional data has shown that the spectra of at least
three pulsars have high-energy turnovers (see Figure 2) which cannot be fit
by a single power-law over the entire EGRET energy range.  Restricting the
spectral fit to the energy range over which the pulsar emission is consistent
with a power law shows that Vela, PSR B1706-44, and PSR B1055-52 have
consistent spectral indices.  In addition, PSR B1951+32 has a softer spectral
index than the younger pulsars Vela and PSR B1706-44.  So there does not seem
to be a correlation between spectral hardness and characteristic age.

It is noteworthy that at least five of the high-energy gamma-ray pulsars
have light curves exhibiting more than one peak, and the sixth pulsar has
insufficient statistics to resolve the pulse morphology.  This suggests that
multiple-peak pulse profiles are a natural consequence of gamma-ray emission
from pulsars.

Double-peaked emission patterns are not unexpected in current models of gamma
ray production from pulsars (e.g. Daugherty \& Harding \cite{daugherty};
Dermer \& Sturner \cite{dermer}; Yadigaroglu \& Romani \cite{yad}).  Models
should now also address the strong bridge emission---as well as the sizeable
off-pulse emission---evident from the three brightest gamma-ray pulsars.

\begin{acknowledgements}
This work was supported by NASA grant NAG5-1605 at Stanford, NASA
grant NAG5-1742 at HSC, NASA contract NAS5-31210 at Grumman, and
Bundesministerium f\"ur Forschung und Technologie grant 50~QV~9095
at MPE.
\end{acknowledgements}

\end{document}